\documentclass[intlimits,twoside,a4paper]{article}

\usepackage{amsmath,amssymb}
\usepackage{graphicx}
\usepackage{wrapfig}
\usepackage{siunitx}
\usepackage{cite}

\usepackage[T2A]{fontenc}
\usepackage[cp1251]{inputenc}

\usepackage[eqsecnum]{cmpj2}

\issue{2012}{15}{4}{43008}

\doinumber{10.5488/CMP.15.43008}

\begin{document}

\title[Simple flexible polymers in a spherical cage]{Simple flexible polymers in a spherical cage}

\author[M. Marenz \textsl{et al.}]{M.~Marenz\refaddr{ITPAdress}, J.~Zierenberg\refaddr{ITPAdress}, H.~Ark{\i}n\refaddr{ITPAdress,Ankara}, W.~Janke\refaddr{ITPAdress}}
\addresses{
\addr{ITPAdress} Institut f\"{u}r Theoretische Physik, Universit\"{a}t Leipzig, Postfach 100920, 04009 Leipzig, Germany
\addr{Ankara} Department of Physics Engineering, Faculty of Engineering, Ankara University, Tandogan, 06100 Ankara, Turkey
}

\authorcopyright{M.~Marenz, J.~Zierenberg, H.~Ark{\i}n, W.~Janke, 2012}

\date{Received July 2, 2012, in final form November 13, 2012}
\maketitle

\begin{abstract}
 We report the results of Monte Carlo simulations investigating the effect of a spherical confinement
 within a simple model for a flexible homopolymer.
 We use the parallel tempering method combined with multi-histogram reweighting analysis and
 multicanonical simulations to investigate thermodynamical observables over a broad range of temperatures,
 which enables us to describe the behavior of the polymer and to locate the freezing and collapse transitions.
 We find a strong effect of the spherical confinement on the location of the collapse transition,
 whereas the freezing transition is hardly effected.

\end{abstract}

\section{Introduction}

The behavior of proteins in confinements has been studied in theory and experiments for a while.
It is a rewarding topic for research because crowded environments
such as caverns, cells, micelles, etc., are the natural habitat of biopolymers,
and the structural conformation has quite an impact on important subjects
such as building biosensors \cite{Service1995},
packaging of DNA \cite{deGennes1999} or the folding behavior of proteins~\cite{Klimov2002,Takagi2003,Friedel2003a,Rathore2006a}.
In this work we are looking at the behavior of a polymer captured in a steric sphere,
which can be considered as a simple model for a polymer in a micelle,
chaperonin-like cage or small pore in a synthetic matrix,
without a complex thermodynamic behavior of the confining structure itself.
There have been simulations with G\~o-like protein models
such as $\beta$-barrel or $\beta$-hairpin proteins and some others \cite{Klimov2002,Takagi2003,Friedel2003a,Rathore2006a,Arkin2011}
in similar confinements.
To get a general overview on the effects of the confinement,
we discard the complexity of 20 different amino acids,
which leads to a large variety of realizations for proteins,
or an enormous amount of different building blocks for synthetic polymers.
Instead, we use a simple bead-stick homopolymer model,
which gives a good overview on general characteristics.
As a first approach, we model the sphere as a steric wall without any attractive or repulsive potential.
We monitor the change of the collapse and freezings transitions and their temperatures $T^{\Theta}_{\mathrm{c}}$ and $T^{\mathrm{F}}_{\mathrm{c}}$
induced by the reduction of the translational entropy and the available space due to the sphere compared with the free polymer.
Although it is a relatively simple model,
the energy landscape is complex enough and the density of states ranges over many orders of magnitude.
Thus, advanced Monte Carlo techniques are necessary
to systematically investigate  the thermodynamic behavior of energetic and conformational observables.

The rest of the paper is organized as follows.
In section~\ref{sec:Model} we describe the used model and observables in detail,
and in section~\ref{sec:SimulationMethods} we briefly review the parallel tempering and multicanonical simulation methods.
Afterwards in section~\ref{sec:Results} we present our results,
and in the last section~\ref{sec:Conclusion} we give a short conclusion.

\section{Model}
\label{sec:Model}

The homopolymer model that we use is a specific form of a model for a heteropolymer
which has been used earlier for investigations of protein folding from a mesoscopic perspective~\cite{Stillinger1993,Irback,Bachmann}.
The polymer consists of $N$ identical monomers,
where the $i$th monomer can be found at position $\vec{r}_i$, and the bending angle between two bonds $\theta_i$ is defined
as $\cos \theta_i = ( \vec{r}_{i+1} - \vec{r}_{i} ) \cdot (\vec{r}_{i+2} - \vec{r}_{i+1})$.
As for lattice models, we neglect any bond vibrations, and adjacent monomers are connected via fixed bonds,
the distances between these monomers $|\vec{r}_{i} - \vec{r}_{i+1}|$ being set to unity.
The excluded volume and attractive parts of the monomer-monomer interaction are modeled by a 12--6 Lennard-Jones potential
for all non-adjacent monomers, the stiffness is introduced via a bending potential and
the confinement by suppressing any state
where at least one monomer is located outside a sphere centered around the origin.
Summarizing, the Hamiltonian consists of three terms,
\begin{align}
  \label{form:energyterms}
  H \equiv E = E_{\textrm{LJ}} + E_{\textrm{bend}} + V_{\textrm{sphere}}\,,
\end{align}
with the Lennard-Jones part being of the common 12--6 form
\begin{align}
  \label{form:LJenergy}
  E_{\textrm{LJ}} = 4 \sum\limits_{i=1}^{N-2} \sum\limits_{j=i+2}^{N} \left( \frac{1}{r_{ij}^{12}} - \frac{1}{r_{ij}^{6}} \right),
\end{align}
where $r_{ij} = | \vec{r}_i - \vec{r}_j | $ is the distance between two monomers.
The bending energy is given by the usual cosine potential
\begin{align}
  \label{form:Bendenergy}
  E_{\textrm{bend}} = \kappa \sum\limits_{i=1}^{N-2} \left( 1 - \cos \theta_i \right),
\end{align}
where the parameter $\kappa$ enables us to adjust the stiffness of the polymer.
In the following simulations we set $\kappa$ to $0.25$, thus the polymer is very flexible.
The sphere is modeled via
\begin{align}
  \label{form:Sphere}
  V_{\textrm{sphere}} = \left\{
  \begin{array}[]{l l}
    0 & \quad \text{if all \hspace{0.9mm} $|r_i|  < R_{\mathrm{S}}$}\,, \\
    \infty & \quad \text{if any $|r_i|  \geqslant  R_{\mathrm{S}}$\,, }
  \end{array}
  \right.
\end{align}
where $R_{\mathrm{S}}$ is the radius of the sphere which ranges in our simulation from $2$ to $12$.
The polymer is allowed to move freely inside this sphere.
Too small spheres lead to a completely unphysical behavior,
because the polymer is pressed into conformations smaller than the crystal conformations,
which are reported for a similar model in \cite{Schnabel2009,Schnabel2009-2}, and thus the excluded volume part leads to extremely high energies.
This gives a lower bound for $R_{\mathrm{S}}$,
the upper limit is chosen so that the behavior of the polymer hardly differs from the bulk behavior.

In order to observe the freezing and collapse transition
and to describe the conformational behavior depending on the radius of the sphere, we choose
the following observables.
For the freezing transition, the energetic observables are ideal.
We measure both parts of the energy $E_{\textrm{LJ}}$ and $E_{\textrm{bend}}$ separately
and, of course, the fluctuations of these quantities, $C_v = \frac{\rd}{\rd T}\langle E\rangle $.
Additionally, the squared radius of gyration
$R^2_{\textrm{gyr}}= \sum_{i=1}^{N} ( \vec{r}_i - \vec{r}_{\textrm{cm}} )^2 / N
= \sum_{i=1}^N\sum_{j=1}^N (\vec{r}_i - \vec{r}_j)^2 / 2N^2$
with $\vec{r}_{\textrm{cm}} = \sum_{i=1}^{N} \vec{r}_{i}/N$,
the squared end-to-end distance $R^2_{\textrm{ee}} = | \vec{r}_{1} - \vec{r}_{N}|^2$
and the thermal fluctuations of these,
$\frac{\rd}{\rd T}\langle O\rangle  = \beta^2 \left( \langle OE\rangle  - \langle O\rangle \langle E\rangle  \right)$,
give a good description of the conformational behavior.
The maximum of the heat capacity $C_v$ is a good indicator of the freezing transition,
because at that temperature the polymer moves into a crystal-like structure,
which is associated with a strong energy drop induced by the Lennard-Jones potential.
The maxima of $\frac{\rd}{\rd T}\langle R^{2}_{\textrm{gyr}} \rangle$ and $\frac{\rd}{\rd T}\langle R^2_{\textrm{ee}} \rangle$
are good indicators for the collapse transition,
at which the polymer changes its conformation from an extended form to a globular one.

\section{Simulation methods}
\label{sec:SimulationMethods}

Although we consider a simple polymer model, its phase space is so complex
that the Metropolis Monte Carlo method will lead to misleading results
at low temperatures or near pseudo phase transitions.
We use two advanced Monte Carlo methods to cope with this problem. A recent overview of these problems is given in~\cite{Janke2012}.
The first method is parallel tempering Monte Carlo sampling,
the principle idea is originally described in~\cite{replicaMC,replicaMC2} and the algorithm itself in~\cite{geyerPT,partemp}.
The second method is multicanonical Monte Carlo (MUCA) sampling~\cite{MUCA1,MUCA2}.
We use the first one to get a good overview of a broad temperature range,
and MUCA to check our results especially near the first-order like freezing transition and at low temperatures.
We will briefly introduce both methods here.

\subsection{Parallel tempering method}
\begin{figure}[t]
  \begin{center}
    \centerline{\includegraphics[width=0.8\textwidth]{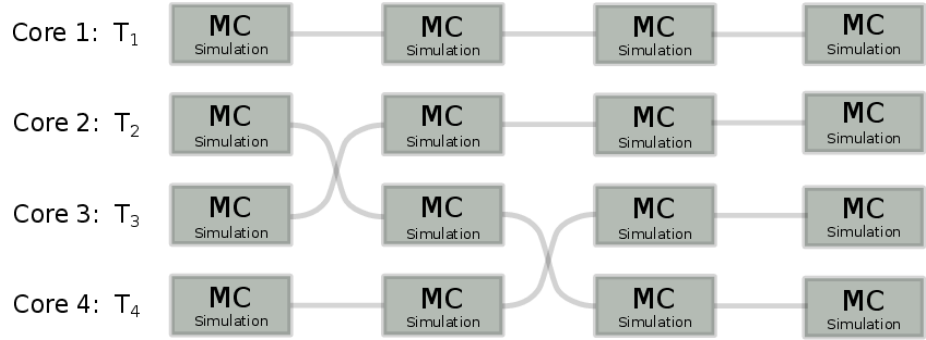}}
  \end{center}
  \caption{A schematic conformation flow between different cores in a parallel tempering simulation.
  Every now and then the conformations are allowed to exchange with probability (\ref{form:pswap}).}
  \label{fig:PT}
\end{figure}

The clue of the parallel tempering method is quite simple.
One runs several separate Metropolis Monte Carlo simulations,
each of them at a different temperature in parallel.
Every now and then two replicas are allowed to exchange their conformation with probability
\begin{align}
  \label{form:pswap}
  p_{\textrm{swap}} = \min\left(1, \re^{\Delta\beta\Delta E}\right),
\end{align}
where $\Delta\beta$ is the difference in the inverse temperature of the replicas
and $\Delta E$ is the difference in the energy of the replicas.
In an implementation of the parallel tempering method, one will not exchange the complete state of the system.
Instead, one would exchange just the temperature and do a little bit bookkeeping to get everything done right.
This procedure is summarized in figure~\ref{fig:PT}.
At the end, the whole simulation yields separate time series for each temperature,
which is a good starting point for the multi-histogram reweighting technique (WHAM) \cite{WHAM1,WHAM2}.
From the individual energy histograms
at every temperature, WHAM calculates the density of states $\Omega\left( E \right)$ in an iterative way.
As a good starting point for this iteration, we use $\Omega\left( E \right)$
obtained by a direct histogram reweighting method~\cite{Fenwick}
which gives a good first estimate for $\Omega\left( E \right)$ and, therefore, leads to a faster convergence.
The parallel tempering  method benefits from the possibility that  a
   single Metropolis Monte Carlo simulation, which is possibly stuck in a
   conformation at low temperature or near a phase transition, can exchange
   its state with a replica from a higher temperature and thereby overcome  its
   stuck state.
This exchange is only possible if the energy distributions of the two temperatures have a sufficient overlap,
which means that the temperature difference between two neighboring replicas should be small enough.
This is also the reason for the weakness of the parallel tempering method
at low temperatures and at first-order phase transitions.
At low temperatures, the energy histograms become very narrow
and one needs many different replicas to cover a broad temperature range.
At first-order phase transitions, the energy distribution is double peaked
with an extremely suppressed regime between the peaks,
so that the parallel tempering method still suffers here from the weakness of the Metropolis Monte Carlo method
which has the problem of overcoming this extremely suppressed region,
but MUCA is capable of countering exactly this problem.

\subsection{Multicanonical Monte Carlo sampling}

The multicanonical method allows one to use arbitrary configuration weights instead of Boltzmann weights  to sample the  phase space of the system.
Therefore, the canonical partition function is modified:
\begin{eqnarray*}
  \label{form:ZMUCA}
  Z_{\textrm{can}} = \sum\limits_{E} \Omega\left( E \right) \re^{-\beta E} &
  \longrightarrow & Z_{\textrm{MUCA}} = \sum\limits_{E}\Omega\left( E \right) W\left( E \right).
\end{eqnarray*}
With this modification one can try to adjust the weights $W\left( E \right)$ in such a way
that the simulation spends equal amounts of time at each energy.
To obtain this, the configuration weight should be equal
to the inverse density of states: $W\left( E \right) = \Omega^{-1}\left( E \right)$.
The density of states is naturally unknown before the simulation.
Therefore, the weights should be  somehow calculated during the simulation.
A possible way to do this is by iteration.
The simplest approach is to start at arbitrary weights, run a simulation with this weight,
calculate the energy distribution $H(E)$
and modify the weights via $W^{(n+1)}(E) = W^{(n)}(E)/H^{(n)}(E)$.
This procedure is repeated until the resulting histograms are flat and span the desired energy range.
At the end, one can reweight the result from an equilibrium production run
with the last weights to every temperature whose Boltzmann energy distribution
lies within the flat energy histogram.
A possible method for this is time-series reweighting,
where every measured observable is weighted
by $W^{-1}\left( E \right) \re^{-\beta E}$ which results in the following formula:
\begin{align}
  \label{form:TSReweighting}
  \left\langle  O \right\rangle _{\beta} =
  \frac{\left\langle  O_i W^{-1}\left( E_i \right) \re^{-\beta E_i } \right\rangle _{\textrm{MUCA}}}
  {\left\langle  W^{-1}\left( E_i \right) \re^{-\beta E_i } \right\rangle _{\textrm{MUCA}}}\,.
\end{align}

\section{Results}
\label{sec:Results}

\begin{figure}[!t]
  \begin{center}
\centerline{\includegraphics[width=0.5\textwidth]{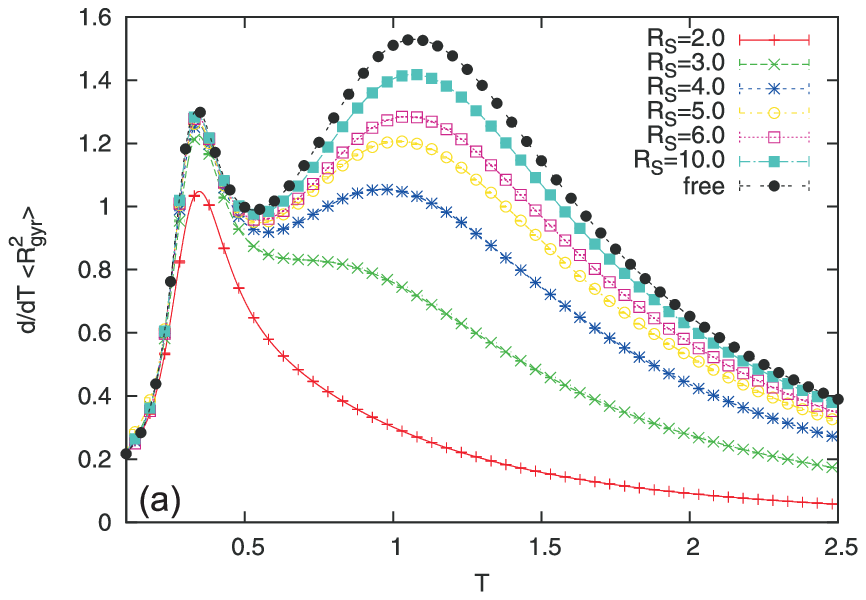}
        \includegraphics[width=0.5\textwidth]{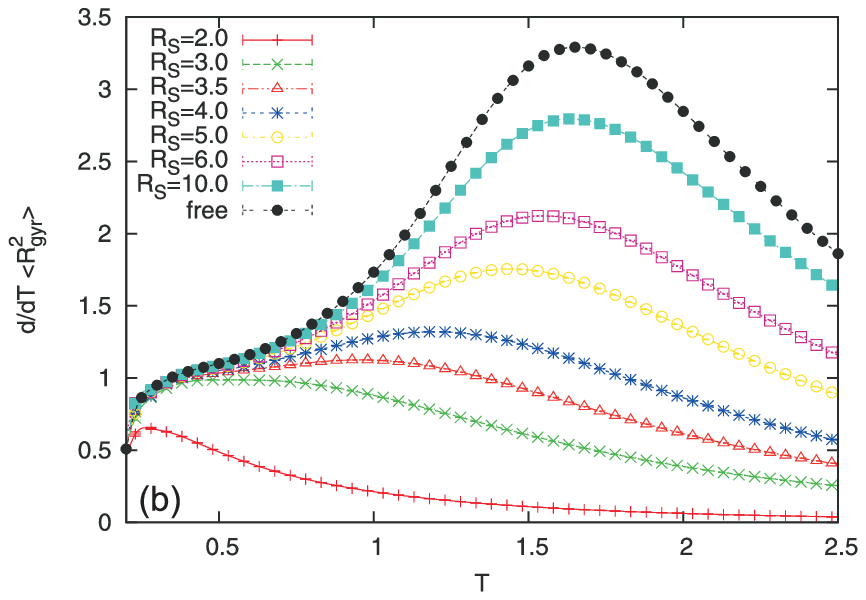}}
    \vspace{5mm}
    \centerline{\includegraphics[width=0.5\textwidth]{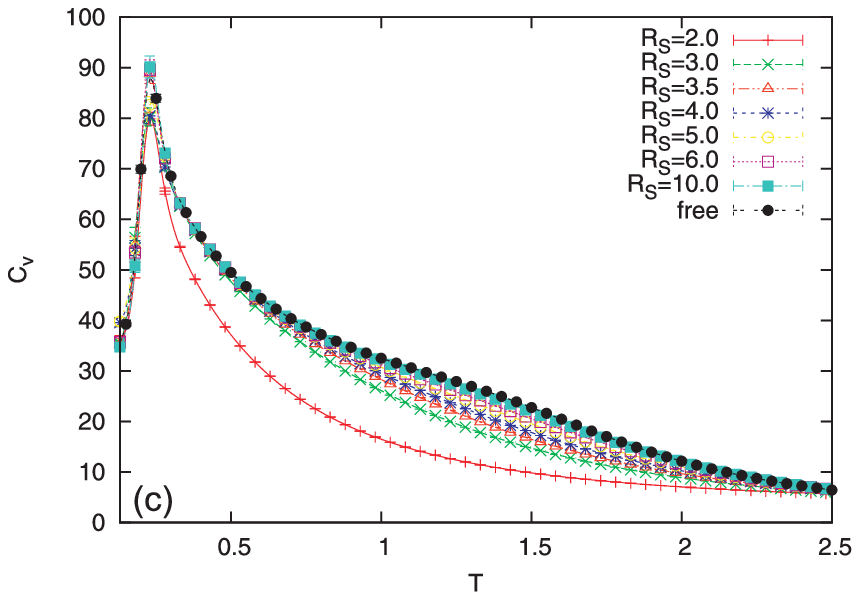}}
  \end{center}
\vspace{-5mm}
  \caption{(Color online) (a) Temperature derivative of $\langle R^2_{\textrm{gyr}} \rangle$ for a 14mer.
(b) Temperature derivative of $\langle R^2_{\textrm{gyr}} \rangle$ for a 28mer.
  (c) Heat capacity for a 28mer.
    For every plot, a subset of the simulated radii is shown.
    The values are calculated with WHAM over the complete temperature range,
    the statistical errors are calculated with the jackknife blocking method and displayed for a sample of all values,
    usually the errors are of the order of the line width.}
  \label{fig:dT_Overview}
\end{figure}

\begin{figure}[!b]
  \begin{center}
\centerline{\includegraphics[width=0.6\textwidth]{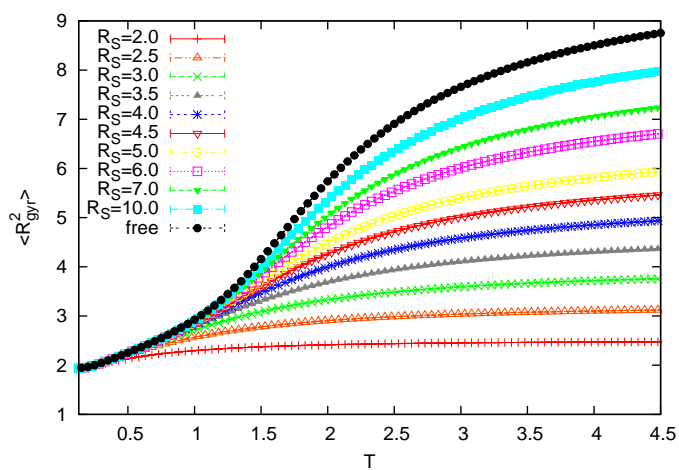}}
  \end{center}
\vspace{-5mm}
  \caption{(Color online) Change of the radius of gyration due a decreasing radius of the sphere.
  Plotted is $\langle R_{\textrm{gyr}}^2 \rangle$ versus the temperature for different sphere radii $R_{\mathrm{S}}$ $(R_{\mathrm{S}}= \text{free},10,\dots,2)$.}
  \label{fig:RadOfGyr_28}
\end{figure}

To locate the pseudo phase transition, we consider the maxima of the temperature derivative of $E$, $R_{\textrm{gyr}}^2$ and $R_{\textrm{ee}}^2$.
For short chain lengths, one can see both pseudo phase transitions
in the temperature-derivative of $\langle  R_{\textrm{gyr}}^2 \rangle $. For longer chains, the peaks for the freezing transition are suppressed,
but still visible in the heat capacity $C_v$, see figure~\ref{fig:dT_Overview}.
The qualitative behavior of $\frac{\rd}{\rd T} \langle R_{\textrm{ee}}^2 \rangle$
is the same as that of $\frac{\rd}{\rd T}\langle R_{\textrm{gyr}}^2 \rangle$.
The effects of the sphere on the elongation  of the polymer are easily predictable.
In the extended phase, above the collapse transition, the extension of the polymer is clearly limited by the sphere.
This effect is still visible but reduced in the collapsed phase, between the freezing and collapse transition,
and hardly visible in the frozen phase, see figure~\ref{fig:RadOfGyr_28}.

In figure~\ref{fig:dT_Overview}, we see that $T_{\textrm{max}}^{\mathrm{F},N}$,
which denotes the temperature of the maximum of the peak at the freezing transition for a fixed $N$,
and the width of the peaks remain similar for different $R_{\mathrm{S}}$,
except for very small $R_{\mathrm{S}}$ where the polymer is pressed into very narrow states.
For these small spheres, the collapse transition vanishes completely,
neither the peak of $\frac{\rd}{\rd T} \langle R_{\textrm{gyr}}^{2} \rangle$ nor the shoulder in $C_v$ exists.
It depends on the length of the polymer at which $R_{\mathrm{S}}$ this effect takes place.

\begin{figure}[h]
  \begin{center}
\centerline{\includegraphics[width=0.5\textwidth]{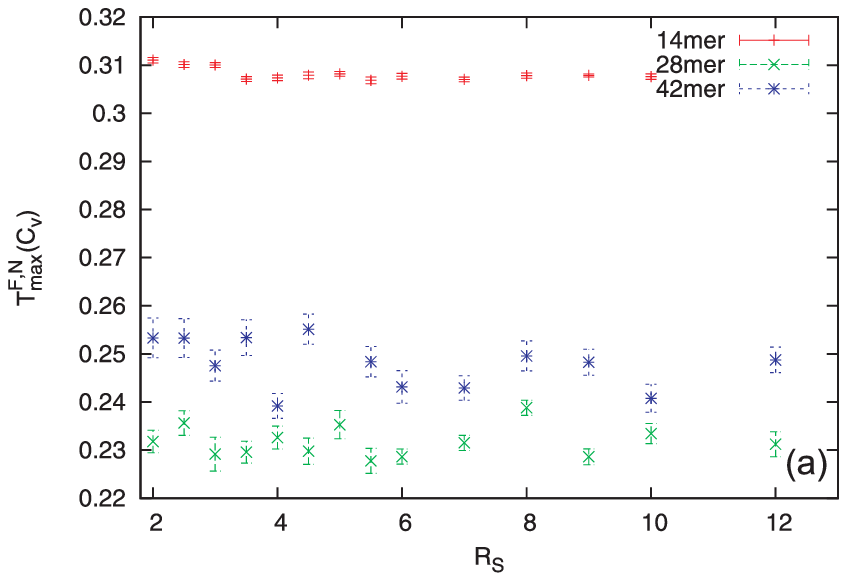}
\includegraphics[width=0.5\textwidth]{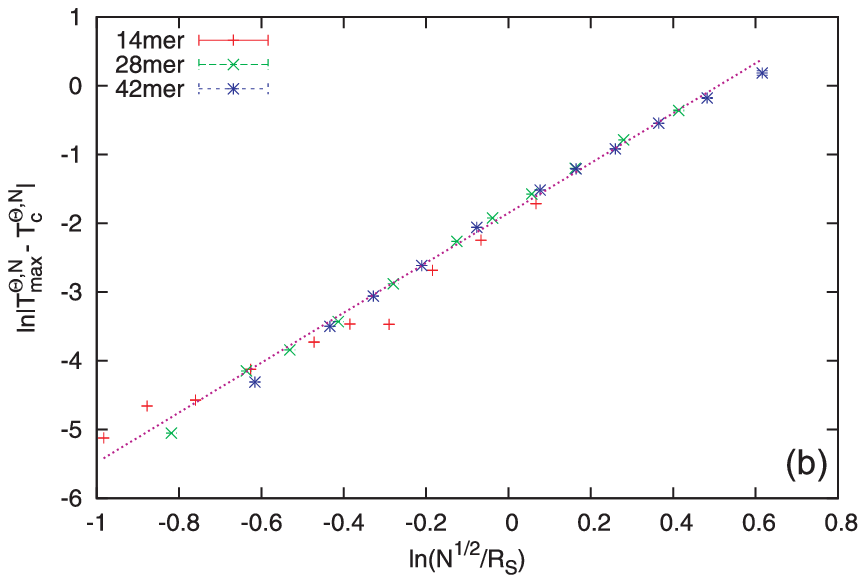}}
  \end{center}
  \vspace{-5mm}
  \caption{(Color online) (a) Temperature of the freezing transition $T_{\textrm{max}}^{\mathrm{F},N}$ versus $R_{\mathrm{S}}$ for different lengths of the polymer $N$ $(N=14,28,42)$.
  One can hardly see any effect of $R_{\mathrm{S}}$ on the location of the freezing transition.
  (b) The change in the collapse transition $|T_{\textrm{max}}^{\Theta,N} - T_{\mathrm{c}}^{\Theta,N}|$
  against $N^{{1}/{2}}/R_{\mathrm{S}}$ for different polymer lengths on a log-log scale.}
  \label{fig:Tc}
\end{figure}

On the other hand, the collapse transition and its temperature $T_{\textrm{max}}^{\Theta,N}$ are strongly effected by the confinement.
The peaks of $\frac{\rd}{\rd T} \langle R_{\textrm{gyr}}^{2} \rangle$ decrease, become broader and shift to lower temperatures as $R_{\mathrm{S}}$ decreases.
A decreasing radius of the sphere pushes the polymer into more collapsed conformations even above the collapse transition.
Thus, the difference in conformational observables between the collapsed and the extended phase decreases, which explains the broader and lower peaks.
The direction of the shift of $T_{\textrm{max}}^{\Theta,N}$ is opposite to the behavior of models for proteins
reported in \cite{Klimov2002,Takagi2003,Friedel2003a,Arkin2011,Rathore2006a},
where the folding temperature increases with a decrease of the available space.
We have simulated a flexible polymer,
whereas these works handle relative short proteins which are more in the semi-flexible or stiff regime.
The different stiffness is a possible explanation for the different behavior.
To quantitatively study  the shift of $T_{\textrm{max}}^{\Theta,N}$  we plot $|T_{\textrm{max}}^{\Theta,N} - T_{\mathrm{c}}^{\Theta,N}|$
versus $N^{\frac{1}{2}}/R_{\mathrm{S}}$ in the log-log plot of figure~\ref{fig:Tc} (right).
Here, $T_{\mathrm{c}}^{\Theta,N}$ denotes the peak location for a free polymer of length $N$.
First, we observe that by using the scaling variable $N^{\frac{1}{2}}/R_{\mathrm{S}}$,
the data for different chain length $N$ and sphere radii $R_{\mathrm{S}}$ fall indeed onto a common master curve.
A linear regression of these data points leads to the scaling behavior
\begin{align}
  \label{form:scaling}
  |T_{\textrm{max}}^{\Theta,N} - T_{\mathrm{c}}^{\Theta,N}| \propto \left( \frac{N^{\frac{1}{2}}}{R_{\mathrm{S}}} \right)^{3.63(15)}.
\end{align}
A similar scaling $|T_{\textrm{max}}^{\Theta,N}  - T_{\mathrm{c}}^{\Theta,N}| \propto \left( R_0/L \right)^{3.25} $,
where $R_0$ is the size of the polymer and $L$ the length of a confining cylinder,
has been reported in~\cite{Takagi2003} for certain protein models.
In our case,  $R_0 \propto N^{\nu}$ holds with $\nu = {1}/{2}$ as for a random walk:
At the collapse transition, the polymer becomes extended and thus does not ``feel'' the self-avoidance,
and in three dimensions it effectively acts as a random walker where $\nu = {1}/{2}$ (up to logarithmic corrections).

\section{Conclusion}
\label{sec:Conclusion}

We have presented a Monte Carlo study of the effects on the pseudo phase transitions of a
flexible polymer caused by a steric confinement.
Advanced Monte Carlo techniques are used to get a detailed estimate for the heat capacity and radius of gyration and
its temperature derivative.
It is found that the confinement has hardly any effect on the behavior and the location of the freezing transition,
whereas for the collapse transition, the behavior and the location change significantly.
Due to the loss of translational entropy and the reduction of possible extended states, the transition becomes less and less pronounced
with a decreasing radius of the confining sphere.
We found a scaling law for the shift of the location of the collapse transition, which holds for all simulated polymer lengths.
This shift is directed towards lower temperatures with decreasing radius of the sphere,
which is opposite to what has been claimed in other works simulating more realistic models of proteins.
One possible reason is that these proteins are much stiffer than the polymer we simulated in this work.
An intended future task is to find out where this difference comes exactly from.

\section*{Acknowledgements}
This work is partially supported by the European Union and the Free State of Saxony through the ``S{\"a}chsische AufbauBank''
and the Alexander von Humboldt Foundation through a Fellowship for Experienced Researchers.
It is also supported by the DFG (German Science Foundation) through SFB/TRR 102 (project B04), Grant No. JA 483/24--3,
and the Graduate School GSC 185 ``BuildMoNa''.
We also thank the Forschungszentrum J{\"u}lich for time on their supercomputer JUROPA under Grant No. hlz17.

\ukrainianpart

\title{Прості гнучкі полімери у сферичній порожнині}
\author{М. Маренц\refaddr{ITPAdress}, Й. Ціренберг\refaddr{ITPAdress}, Г. Аркін\refaddr{ITPAdress,Ankara}, В. Янке\refaddr{ITPAdress}}
\addresses{
\addr{ITPAdress} Інститут теоретичної фізики, Університет  Ляпцігу, 04009 Ляйпціг, Німеччина
\addr{Ankara} Кафедра фізичної інженерії, інженерний факультет, Університет  Анкари,  06100 Анкара, Туреччина
}

\makeukrtitle

\begin{abstract}
\tolerance=3000%
Ми повідомляємо результати симуляцій Монте Карло  дослідження впливу сферичного просторового обмеження
в рамках простої моделі гнучкого гомополімера. Ми використовуємо метод паралельного темперування в поєднанні з
аналізом методом мультигістограмного перезважування, а також великоканонічні симуляції для того, щоб дослідити
термодинамічні спостережувальні  в широкій області температури, що дозволяє нам описати поведінку і місце знаходження
переходів замерзання і колапсу. Ми виявляємо сильний вплив сферичного просторового обмеження на місце знаходження
переходу колапсу, тоді як перехід замерзання навряд чи піддається впливу.
\end{abstract}

\end{document}